\documentclass[onecolumn, a4size, 12pt]{IEEEtran}
\usepackage{amsmath}
\usepackage{amssymb}
\usepackage{amsfonts}
\usepackage{graphicx}
\usepackage{epsfig}
\usepackage{subfigure}
\usepackage{psfrag}

\title{Optimal Power Control over Fading Cognitive Radio Channels by Exploiting Primary User
CSI}

\author{Rui Zhang \footnote{Rui Zhang is with the Institute for Infocomm Research, A*STAR, Singapore
(e-mail: rzhang@i2r.a-star.edu.sg).}\vspace{-0.4in}}

\begin{document}
\maketitle \maketitle \thispagestyle{empty}

\begin{abstract}
This paper is concerned with spectrum sharing cognitive radio
networks, where a secondary user (SU) or cognitive radio link
communicates simultaneously over the same frequency band with an
existing primary user (PU) link. It is assumed that the SU
transmitter has the perfect channel state information (CSI) on the
fading channels from SU transmitter to both PU and SU receivers (as
usually assumed in the literature), as well as the fading channel
from PU transmitter to PU receiver (a new assumption). With the
additional PU CSI, we study the optimal power control for the SU
over different fading states to maximize the SU ergodic capacity
subject to a new proposed constraint to protect the PU transmission,
which limits the maximum ergodic capacity loss of the PU resulted
from the SU transmission. It is shown that the proposed SU
power-control policy is superior over the conventional policy under
the constraint on the maximum tolerable interference
power/interperferecne temperature at the PU receiver, in terms of
the achievable ergodic capacities of both PU and SU.
\end{abstract}


\setlength{\baselineskip}{1.3\baselineskip}

\newtheorem{theorem}{Theorem}[section]
\newtheorem{remark}{Remark}[section]

\newcommand{\mv}[1]{\mbox{\boldmath{$ #1 $}}}

\section{Introduction}
In this paper, we are concerned with the newly emerging {\it
cognitive radio} (CR) type of wireless communication networks, where
the secondary users (SUs) or the so-called CRs communicate over the
same frequency band that has been allocated to the existing primary
users (PUs). For such scenarios, the SUs usually need to deal with a
fundamental tradeoff between maximizing the secondary network
throughput and minimizing the performance loss of the primary
network resulted from the SU transmissions. One commonly known
technique used by the SUs to protect the PUs is {\it opportunistic
spectrum access} (OSA) \cite{Mitola00}, whereby the SUs decide to
transmit over the channel of interest only if the PU transmissions
are detected to be off. Many algorithms have been reported in the
literature for detecting the PU transmission status, in general
known as {\it spectrum sensing} (see, e.g.,
\cite{Cabric04}-\cite{Sahai06} and references therein). In contrast
to OSA, another general operation model of CRs is known as {\it
spectrum sharing} (SS) \cite{Zhao07}, where the SUs are allowed to
transmit {\it simultaneously} with the PUs provided that the
interferences from the SUs to PUs will cause the resultant PU
performance loss to be within an acceptable level. Under the SS
paradigm, the ``cognitive relay'' idea has been proposed in
\cite{Tarokh06}, \cite{Viswanath06}, where the SU transmitter is
assumed to know {\it a priori} the PU's messages and is thus able to
compensate for the interferences to the PU receiver resulted from
the SU transmission by operating as an assisting relay to the PU
transmission. As an alternative method for SS, the SU can protect
the PU transmission by regulating the SU to PU interference power
level to be below a predefined threshold, known as the {\it
interference temperature} (IT), \cite{Haykin05}, \cite{Gastpar07}.
This method is perhaps more practical than the cognitive relay
method since only the SU to PU channel gain knowledge is required to
be known to the SU transmitter.

In this paper, we focus our study on the SS (as opposed to OSA)
-based CR networks. It is known that in wireless networks, {\it
dynamic resource allocation} (DRA) whereby the transmit powers,
bit-rates, bandwidths and/or antenna beams of users are dynamically
allocated based upon the channel state information (CSI) is
essential to the achievable network throughput. For the case of
single-antenna PU and SU channels, adaptive power control for the SU
is an effective means of DRA, and has been studied in, e.g.,
\cite{Ghasemi07}, \cite{Syed07}, to maximize the SU's transmission
rate subject to the constraint on the maximum tolerable average IT
level at the PU receiver. In \cite{Zhang08}, the authors proposed
both optimal and suboptimal spatial adaptation schemes for the SUs
equipped with multiple antennas under the given IT constraints at
different PU receivers. DRA in the multiuser CR networks based on
the principle of IT has also been studied in, e.g., \cite{Huang05},
\cite{Zhang08MAC}. It is thus known that most existing works in the
literature on DRA for the SUs are based on the IT idea. Although IT
is a practical method to protect the PU transmission, its optimality
on the achievable performance tradeoff for the SU has not yet been
carefully addressed in the literature, to the author's best
knowledge.

In this paper, we consider a simplified fading CR network where only
one pair of SU and PU is present and all the terminals involved are
equipped with a single antenna. We assume that not only the CSI on
the SU fading channel and that from the SU transmitter to PU
receiver is known to the SU (as usually assumed in the literature),
but is also the CSI on the PU fading channel (a new assumption made
in this paper). In practice, CSI on the SU's own channel can be
obtained via the classical channel training and feedback methods,
while CSI from the SU transmitter to PU receiver can be obtained by
the SU transmitter via estimating the reversed channel from PU
receiver under the assumption of channel reciprocity. The more
challenging task is perhaps on obtaining the CSI on the PU fading
channel by the SU, for which more sophisticated techniques are
required, e.g., via eavesdropping the feedback from the PU receiver
to PU transmitter \cite{Gastpar07b}, or via a cooperative secondary
node located in the vicinity of the PU receiver \cite{Li05}. With
the additional PU CSI, the SU is able to transmit with large powers
when the PU fading channel is in inferior conditions such as deep
fading, since under such circumstances the resultant PU performance
loss is negligible, nearly independent of the exact interference
level at the PU receiver. In contrast, for the case where the IT
constraint is applied, the SU's transmit power is determined by the
channel gain from the SU transmitter to PU receiver, while it is
independent of the PU CSI. Motivated by these observations, this
papershows that there in fact exists a better means to protect the
PU transmission as well as to maximize the SU transmission rate than
the conventional IT constraint or more specifically, the {\it
average interference power constraint} (AIPC) over the fading states
at the PU receiver. This new proposed method ensures that the
maximum ergodic capacity loss of the PU resulted from the SU
transmission is no greater than some prescribed threshold, thus
named as {\it primary capacity loss constraint} (PCLC). Clearly, the
PCLC is more directly related to the PU transmission than the AIPC.
In this paper, we will formally study the optimal power-control
policy for the SU under the new proposed PCLC, and show its
performance gains over the AIPC in terms of improved ergodic
capacities of both PU and SU.

The rest of this paperis organized as follows. Section
\ref{sec:channel model} presents the system model. Section
\ref{sec:primary receiver power constraint} presents the
conventional power-control policy for the SU based on the AIPC.
Section \ref{sec:primary capacity loss constraint} introduces the
new PCLC, and derives the associated optimal power-control policy
for the SU. Section \ref{sec:simulation results} provides numerical
examples to evaluate the achievable rates of the proposed scheme.
Finally, Section \ref{sec:conclusion} gives the concluding remarks.

{\it Notation}: $|\cdot|$ denotes the Euclidean norm of complex
number. $\mathbb{E}[\cdot]$ denotes the statistical expectation. The
distribution of a circularly symmetric complex Gaussian (CSCG)
random variable (RV) with mean $x$ and variance $y$ is denoted by
$\mathcal{CN}(x,y)$, and $\sim$ means ``distributed as''.
$(\cdot)^+=\max(0,\cdot)$. The notations $|\cdot|$, $(\cdot)^{T}$,
and $(\cdot)^{H}$ denote the matrix determinant, transpose, and
conjugate transpose, respectively. $\mv{I}$ denotes an identity
matrix. $\|\cdot\|$ denotes the Euclidean norm of complex vector.

\section{System Model} \label{sec:channel model}

\begin{figure}
\psfrag{a}{\small \hspace{1cm} PU-Tx}\psfrag{b}{\small \hspace{1cm}
SU-Tx} \psfrag{c}{\small PU-Rx}\psfrag{d}{\small SU-Rx}
\psfrag{e}{$\tilde{\mv{f}}$}\psfrag{f}{$\tilde{\mv{e}}$}\psfrag{g}{$\tilde{\mv{o}}$}\psfrag{h}{$\tilde{\mv{g}}$}
\begin{center}
\scalebox{1.2}{\includegraphics*[88pt,617pt][302pt,755pt]{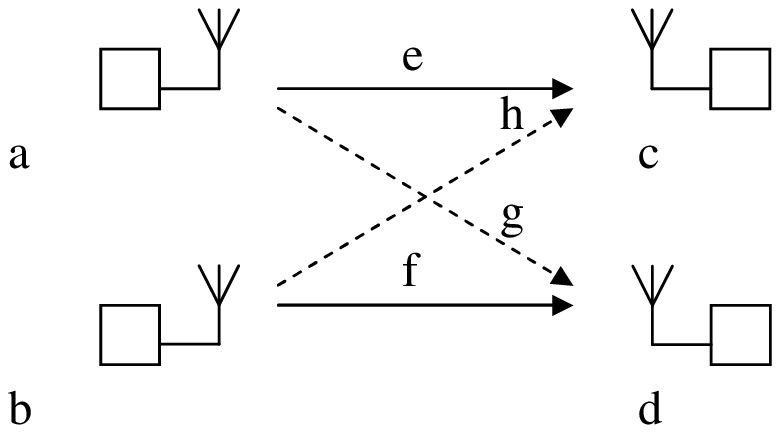}}
\end{center}
\caption{System model for the CR network.}\label{fig:system model}
\end{figure}

As shown in Fig. \ref{fig:system model}, we consider a SS-based CR
network where a SU link consisting of a SU transmitter (SU-Tx) and a
SU receiver (SU-Rx) transmits simultaneously over the same narrow
band with a PU link consisting of a PU transmitter (PU-Tx) and a PU
receiver (PU-Rx). The fading channel complex gains from SU-Tx to
SU-Rx and PU-Rx are represented by $\tilde{\mv{e}}$ and
$\tilde{\mv{g}}$, respectively, and from PU-Tx to PU-Rx and PU-Rx by
$\tilde{\mv{f}}$ and $\tilde{\mv{o}}$, respectively. We assume a
block-fading (BF) channel model and denote $i$ as the joint fading
state for all the channels involved. Furthermore, we assume coherent
communications for both the PU and SU links and thus only the fading
channel power gains (amplitude squares) are of interest. Let $e_i$
denote the power gain of $\tilde{\mv{e}}$ at fading state $i$, i.e.,
$e_i=|\tilde{\mv{e}}(i)|^2$; similarly, $g_i$, $f_i$, and $o_i$ are
defined. It is assumed that $e_i$, $g_i$, $f_i$, and $o_i$ are
independent RVs each having a continuous probability density
function (PDF). It is also assumed that the additive noises at both
PU-Rx and SU-Rx are independent CSCG RVs each
$\sim\mathcal{CN}(0,1)$. Since we are interested in the
information-theoretic limits, it is assumed that the optimal
Gaussian codebook is used by both the PU and SU.

Consider first the PU link. Since the PU may adopt an adaptive power
control based on its own CSI, the PU's transmit power at fading
state $i$ is denoted by $q_i$. It is assumed that the PU's
power-control policy, $\mathcal{P}_p(f_i)$, is a mapping from the PU
channel power gain $f_i$ to $q_i$, subject to an average transmit
power constraint represented by $\mathbb{E}[q_i]\leq Q$. Examples of
PU power control are the ``constant-power (CP)'' policy
\begin{equation}\label{eq:CP}
q_i=Q, \ \forall i
\end{equation}
and the ``water-filling (WF)'' policy \cite{Goldsmith97},
\cite{Coverbook}
\begin{equation}\label{eq:WF}
q_i=\left(d_p-\frac{1}{f_i}\right)^+
\end{equation}
where $d_p$ is a constant ``water-level'' with which
$\mathbb{E}[q_i]=Q$. In this paper, we assume that the PU is
oblivious to the existence of the SU, and any interference from
SU-Tx is treated as additional Gaussian noise at PU-Rx. Thus, the
ergodic capacity of the PU channel is expressed as
\begin{equation}\label{eq:primary capacity}
C_p=\mathbb{E}\left[\log\left(1+\frac{f_iq_i}{1+g_ip_i}\right)\right]
\end{equation}
where $p_i$ denotes the SU's transmit power at fading state $i$ (to
be more specified later). It is worth noting that the maximum PU
ergodic capacity, denoted as $C_p^{\rm
max}=\mathbb{E}\left[\log\left(1+f_iq_i\right)\right]$, is
achievable only if
\begin{equation}\label{eq:max primary capacity condition}
\frac{f_iq_i}{1+g_ip_i}=f_iq_i, \ \forall i.
\end{equation}
From (\ref{eq:max primary capacity condition}), it follows that
$g_ip_i=0$ if $f_iq_i>0$ for any $i$. In other words, to achieve
$C_p^{\rm max}$ for the PU, the SU transmission must be off when the
PU transmission is on, which is the same as the OSA with perfect
spectrum sensing.

Next, consider the SU link. The SU is also known as CR since it is
aware of the PU transmission and is able to adapt its transmit power
levels at different fading states based on all the available CSI
between the PU and SU to maximize the SU's average transmit rate and
yet provide a sufficient protection to the PU. In this paper, we
assume that the CSI on both $g_i$ and $f_i$ is perfectly known at
SU-Tx for each fading state $i$. For notational convenience, we
combine the Gaussian-distributed interference from PU-Tx with the
additive noise at SU-Rx, and define the equivalent SU channel power
gain $\mv{h}$ at fading state $i$ as $h_i:= \frac{e_i}{1+o_iq_i}$,
which is also assumed to be known at SU-Tx for each $i$. Thus, the
SU's power-control policy can be expressed as
$\mathcal{P}_s(h_i,g_i,f_i)$, subject to an average transmit power
constraint, $\mathbb{E}[p_i]\leq P$. By assuming that SU-Rx treats
the interference from PU-Tx as additional Gaussian noise, the SU's
achievable ergodic capacity is then expressed as
\begin{equation}\label{eq:secondary capacity}
C_s=\mathbb{E}\left[\log\left(1+h_ip_i\right)\right].
\end{equation}
Note that the maximum SU ergodic capacity, denoted by $C_s^{\rm
max}$, is achievable when $\mathcal{P}_s$ maximizes $C_s$ with no
attempt to protect the PU transmission. In this case, the optimal
$\mathcal{P}_s$ is the WF policy expressed as
$p_i=\left(d_s-\frac{1}{h_i}\right)^+$, where $d_s$ is the
water-level with which $\mathbb{E}[p_i]=P$.

\begin{remark}
It is important to note that in the assumed system model, we have
deliberately excluded the possibility that the PU's allocated power
at fading state $i$ is a function of the received interference power
from SU-Tx, $g_ip_i$. If not so, the SU's power control needs to
take into account of any predictable reaction of the PU upon
receiving the interference from SU-Tx, e.g., the PU may change
transmit power that will also result in change of the interference
power level at SU-Rx. Such feedback loop over the SU's and PU's
power adaptations will make the design of the SU transmission more
involved even for the deterministic channel case. This interesting
phenomenon will be studied in the future work.
\end{remark}

\section{SU Power Control Under AIPC} \label{sec:primary receiver
power constraint}

Existing prior work in the literature, e.g., \cite{Ghasemi07}, has
considered the peak/average interference power constraint over
fading states at PU-Rx as a practical means to protect the PU
transmission. In this section, we first present the SU power-control
policy to maximize the SU ergodic capacity under the constraint that
the {\it average} interference power level over different fading
states at PU-Rx must be regulated below some predefined threshold,
thus named as {\it average interference power constraint} (AIPC).
The associated problem formulation is similar to that in
\cite{Ghasemi07}, but with an additional constraint on the SU's own
transmit power constraint, and is expressed as
\begin{eqnarray}
\mbox{(P1)}:~~\mathop{\mathtt{Maximize}}_{\{p_i\}}&&
\mathbb{E}\left[\log\left(1+h_ip_i\right)\right] \nonumber
\\ \mathtt {Subject \ to} && \mathbb{E}[g_ip_i]\leq \Gamma \label{eq:IPC} \\
&& \mathbb{E}[p_i]\leq P  \label{eq:TX constraint} \\ && p_i\geq 0,
\ \forall i \label{eq:power positive constraint}
\end{eqnarray}
where $\Gamma\geq 0$ is the predefined threshold for AIPC. It is
easy to verify that (P1) is a convex optimization problem, and thus
by applying the standard Karush-Kuhn-Tucker (KKT) optimality
conditions \cite{Boydbook}, the optimal solution of (P1), denoted as
$\{p_i^{(1)}\}$, is obtained as
\begin{equation}\label{eq:optimal power IPC}
p_i^{(1)}=\left(\frac{1}{\nu^{(1)}
g_i+\mu^{(1)}}-\frac{1}{h_i}\right)^+
\end{equation}
where $\nu^{(1)}$ and $\mu^{(1)}$ are non-negative
constants.\footnote{Numerically, $\nu^{(1)}$ and $\mu^{(1)}$ can be
obtained by, e.g., the ellipsoid method \cite{BGT81}. This method
utilizes the sub-gradients $\Gamma-\mathbb{E}[g_ip_i[n]]$ and
$P-\mathbb{E}[p_i[n]]$ to iteratively update $\nu[n+1]$ and
$\mu[n+1]$ until they converge to $\nu^{(1)}$ and $\mu^{(1)}$,
respectively, where $\{p_i[n]\}$ is obtained from (\ref{eq:optimal
power IPC}) for some given $\nu[n]$ and $\mu[n]$ at the $n$th
iteration.}

Note that the power-control policy given in (\ref{eq:optimal power
IPC}) is a {\it modified} version of the standard WF policy
\cite{Goldsmith97}, \cite{Coverbook}. Compared with the standard WF
policy, (\ref{eq:optimal power IPC}) differs in that the water-level
is no longer a constant, but is instead a function of $g_i$.
Interestingly, similar variable water-level WF power control has
also been shown in \cite{Yu07} for multi-carrier systems in the
presence of multiuser crosstalk. If $g_i=0$, (\ref{eq:optimal power
IPC}) becomes the standard WF policy with a constant water-level
$1/\mu^{(1)}$, since in this case the SU transmission does not
interfere with PU-Rx. On the other hand, if $g_i\rightarrow \infty$,
from (\ref{eq:optimal power IPC}) it follows that the water-level
becomes zero and thus $p_i^{(1)}=0$ regardless of $h_i$, suggesting
that in this case no SU transmission is allowed since any finite SU
transmit power will result in an infinite interference power at
PU-Rx.

Furthermore, it is observed from (\ref{eq:optimal power IPC}) that
the power control under AIPC does not require the PU CSI, $f_i$,
which is desirable from an implementation viewpoint. However, there
are also drawbacks of this power control explained as follows.
Supposing that $f_iq_i=0$, i.e., the PU transmission is off, the SU
can not take this opportunity to transmit if $g_i$ happens to be
sufficiently large such that (\ref{eq:optimal power IPC}) results in
that $p_i^{(1)}=0$. On the other hand, if $f_iq_i$ happens to be a
large value, suggesting that a substantial amount of information is
transmitted over the PU channel, such transmission may be corrupted
by a strong interference from the SU if in (\ref{eq:optimal power
IPC}) $g_i$ and $h_i$ result in a large interference power
$g_ip_i^{(1)}$ (though it is upper-bounded by $\frac{1}{\nu^{(1)}}$)
at PU-Rx. Clearly, the above drawbacks of the AIPC-based power
control are due to the lack of joint exploitation of all the
available CSI at SU-Tx, which will be overcome by the proposed
power-control policy in the next section.

It is worth mentioning that although the AIPC-based power control is
non-optimal, the AIPC still guarantees an upper bound on the maximum
PU ergodic capacity loss, as given by the following theorem:
\begin{theorem}\label{theorem}
Under the given AIPC threshold, $\Gamma$, the PU ergodic capacity
loss due to the SU transmission, defined as $C_p^{\rm max}-C_p$, is
upper-bounded by $\log(1+\Gamma)$, regardless of
$\mathcal{P}_p(f_i)$, $\mathcal{P}_s(h_i,g_i,f_i)$, and the
distributions of $f_i$, $h_i$, and $g_i$.
\end{theorem}
\begin{proof}
The proof is based on the following equality/inequalities:
\begin{align}
C_p
&\overset{(a)}{=}\mathbb{E}\left[\log\left(1+\frac{f_iq_i}{1+g_ip_i}\right)\right]
\nonumber
\\ & \overset{(b)}{\geq}
\mathbb{E}\left[\log\left(1+f_iq_i\right)\right]-\mathbb{E}\left[\log\left(1+g_ip_i\right)\right]
\nonumber
\\ &  \overset{(c)}{\geq}
\mathbb{E}\left[\log\left(1+f_iq_i\right)\right]-\log\left(1+\mathbb{E}[g_ip_i]\right)
\nonumber
\\ & \overset{(d)}{\geq} C_p^{\rm max}-\log(1+\Gamma) \nonumber
\end{align}
where $(a)$ is due to (\ref{eq:primary capacity}); $(b)$ is due to
$g_ip_i\geq 0, \forall i$; $(c)$ is due to the concavity of the
function $\log(1+x)$ for $x\geq0$ and Jensen's inequality (see,
e.g., \cite{Coverbook}); and $(d)$ is due to the definition of
$C_p^{\rm max}$ and the inequality (\ref{eq:IPC}).
\end{proof}

\section{SU Power Control Under PCLC} \label{sec:primary capacity loss constraint}

In this section, we propose a new SU power-control policy by
utilizing all the CSI on $h_i$, $g_i$, and $f_i$, known at SU-Tx.
This new policy is based on an alternative constraint of AIPC to
protect the PU transmission, named as {\it primary capacity loss
constraint} (PCLC). PCLC and AIPC are related to each other: From
Theorem \ref{theorem}, it follows that the AIPC in (\ref{eq:IPC})
with a given $\Gamma$ implies that $C_p^{\rm max}-C_p\leq
\log(1+\Gamma)$, while the PCLC directly applies the constraint
$C_p^{\rm max}-C_p\leq C_{\delta}$, where $C_{\delta}$ is a
predefined value for the maximum tolerable ergodic capacity loss of
the PU resulted from the SU transmission.\footnote{Since most
communication systems in practice employ some form of ``power
margin'' and/or ``rate margin'' (see, e.g., \cite{Cioffi}) for the
receiver to deal with unexpected interferences, the PCLC is a valid
assumption if the PU belongs to such systems.} The ergodic capacity
maximization problem for the SU under the PCLC and the SU's own
transmit power constraint is expressed as
\begin{eqnarray}
\mbox{(P2):}~~\mathop{\mathtt{Maximize}}_{\{p_i\}}&&
\mathbb{E}\left[\log\left(1+h_ip_i\right)\right] \nonumber
\\ \mathtt {Subject \ to} && C_p^{\rm max}-C_p\leq C_{\delta} \label{eq:PCLC} \\
&&  (\ref{eq:TX constraint}), (\ref{eq:power positive constraint}).
\nonumber
\end{eqnarray}
Note that (P1) and (P2) only differ in the constraints,
(\ref{eq:IPC}) and (\ref{eq:PCLC}). Since $C_p^{\rm max}$ is a fixed
value given $Q$, the distribution of $f_i$, and the PU power control
$\mathcal{P}_{p}(f_i)$, using (\ref{eq:primary capacity}) we can
rewrite (\ref{eq:PCLC}) as
\begin{equation}\label{eq:PCLC new}
\mathbb{E}\left[\log\left(1+\frac{f_iq_i}{1+g_ip_i}\right)\right]\geq
C_0
\end{equation}
where $C_0=C_p^{\rm max}-C_{\delta}$. Unfortunately, the constraint
(\ref{eq:PCLC new}) can be shown to be non-convex, rendering (P2) to
be also non-convex. However, under the assumption of continuous
fading channel gain distributions, it can be easily verified that
the so-called ``time-sharing'' condition given in \cite{Yu06} is
satisfied by (P2). Thus, we can solve (P2) by considering its
Lagrange dual problem, and the resultant duality gap between the
original and the dual problems is zero. Due to the lack of space, we
skip here the detailed derivations and present the solution of (P2)
directly as follows:
\begin{equation}\label{eq:optimal power PCLC}
p_i^{(2)}=\left(\frac{1}{\lambda_i(p_i^{(2)})\nu^{(2)}
g_i+\mu^{(2)}}-\frac{1}{h_i}\right)^+
\end{equation}
where similarly like (P1),  $\nu^{(2)}$ and $\mu^{(2)}$ are
nonnegative constants that can be obtained by the ellipsoid method.
Compared to the AIPC-based power-control policy in (\ref{eq:optimal
power IPC}), the new policy in (\ref{eq:optimal power PCLC}) based
on PCLC has an additional multiplication factor in front of the term
$\nu^{(2)}g_i$, which is further expressed as
\begin{equation}\label{eq:lambda i}
\lambda_i\left(p_i^{(2)}\right)=\frac{f_iq_i}{\left(1+g_ip_i^{(2)}\right)\left(1+g_ip^{(2)}_i+f_iq_i\right)}.
\end{equation}
Note that $\lambda_i$ is itself a (decreasing) function of the
optimal solution $p_i^{(2)}$. Thus, the power-control policy
(\ref{eq:optimal power PCLC}) can be considered as a {\it
self-biased} WF solution. From (\ref{eq:optimal power PCLC}) and
(\ref{eq:lambda i}), we obtain
\begin{theorem}\label{theorem:new}
The optimal solution of (P2) is
\begin{eqnarray}\label{eq:optimal power control final}
p_i^{(2)}=\left\{\begin{array}{ll} 0 & {\rm if} \
\frac{1}{\lambda_i(0)\nu^{(2)} g_i+\mu^{(2)}}-\frac{1}{h_i}\leq 0
\\ z_0 & {\rm otherwise,}
\end{array} \right.
\end{eqnarray}
where $z_0$ is the unique positive root of $z$ in the following
equation:
\begin{eqnarray}\label{eq:WF equation}
z=\frac{1}{\lambda_i(z)\nu^{(2)} g_i+\mu^{(2)}}-\frac{1}{h_i}.
\end{eqnarray}
\end{theorem}

\begin{figure}
\psfrag{a}{$F(z)$}\psfrag{b}{$z$}
\psfrag{c}{$\frac{1}{h_i}$}\psfrag{d}{$\frac{1}{\mu^{(2)}}$}
\psfrag{e}{$z_0$}\psfrag{f}{$\measuredangle 45$}
\begin{center}
\scalebox{1.0}{\includegraphics*[61pt,521pt][364pt,766pt]{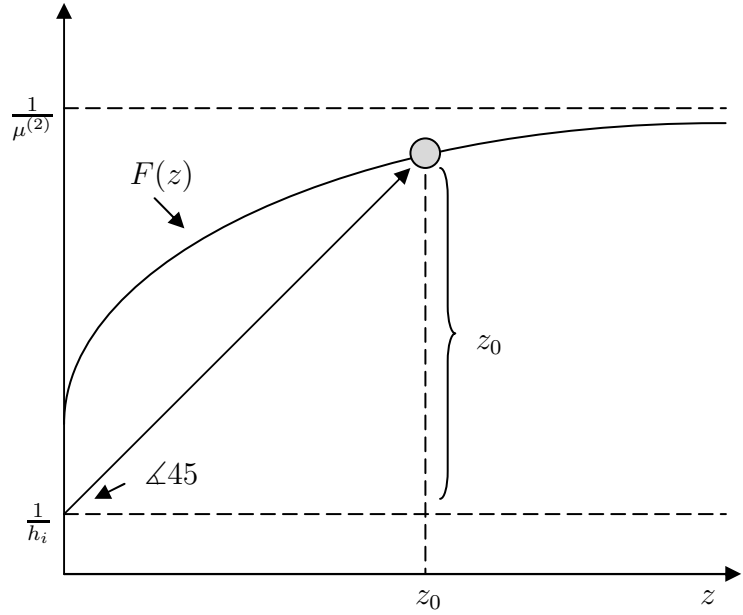}}
\end{center}
\caption{Illustration of the unique positive root $z_0$ for the
equation (\ref{eq:WF equation}).}\label{fig:WF}
\end{figure}

An illustration of the unique positive root $z_0$ for the equation
(\ref{eq:WF equation}) is given in Fig. \ref{fig:WF}. Note that
$F(z)\triangleq \frac{1}{\lambda_i(z)\nu^{(2)} g_i+\mu^{(2)}}$ is an
increasing function of $z$ for $z\geq 0$, and $F(0)\geq
\frac{1}{h_i}$, $F(\infty)=\frac{1}{\mu^{(2)}}$. As shown, $z_0$ is
obtained as the intersection between a 45-degree line starting from
the point $(0,\frac{1}{h_i})$ and the curve showing the values of
$F(z)$. Numerically, $z_0$ can be obtained by a simple bisection
search \cite{Boydbook}.

Some interesting observations are drawn on the PCLC-based power
control (\ref{eq:optimal power control final}) as follows:

First, from (\ref{eq:optimal power PCLC}) and (\ref{eq:lambda i}) it
is observed that what is indeed required at SU-Tx for power control
at each fading state is the received signal power at PU-Rx,
$f_iq_i$, instead of the exact PU channel CSI, $f_i$. Note that
$f_iq_i$ may be more easily obtainable by the SU than $f_i$ in some
cases, e.g., when $f_iq_i$ is estimated and then sent back by a
collaborate secondary node in the vicinity of PU-Rx.

Second, from (\ref{eq:lambda i}) and (\ref{eq:optimal power control
final}) it is inferred that $p_i^{(2)}>0$ for any $i$ if and only if
$\frac{f_iq_i}{1+f_iq_i}\nu^{(2)}g_i+\mu^{(2)}<h_i$. For given
$\nu^{(2)}$, $\mu^{(2)}$, and $h_i$, it then follows that
$p_i^{(2)}>0$  only when $g_i$ and/or $f_iq_i$ are sufficiently
small. This is intuitively correct because they are indeed the cases
where the SU will cause only a negligible PU capacity loss. In the
extreme case of $g_i=0$ and/or $f_iq_i=0$, the condition for
$p_i^{(2)}>0$ becomes $\frac{1}{\mu^{(2)}}>\frac{1}{h_i}$, the same
as the standard WF policy.

\section{Numerical Examples}\label{sec:simulation
results}

The achievable ergodic capacity pairs of PU and SU, denoted by
$(C_p,C_s)$, over realistic fading channels are presented in this
section via simulation. $\tilde{\mv{f}}$, $\tilde{\mv{e}}$,
$\tilde{\mv{g}}$, and $\tilde{\mv{o}}$ are assumed to independent
CSCG RVs $\sim$ $\mathcal{CN}(0,1)$, $\mathcal{CN}(0,1)$,
$\mathcal{CN}(0,0.5)$, and $\mathcal{CN}(0,0.01)$, respectively. It
is also assumed that $P=Q=10$, corresponding to an equivalent
average signal-to-noise ratio (SNR) of 10 dB for both PU and SU
channels (without the interference between PU and SU). The following
cases of $(C_p,C_s)$ are then considered:
\begin{itemize}
\item ``PCLC'': The SU employs the proposed power-control policy (\ref{eq:optimal power control
final}).
 $C_s$'s are obtained from (\ref{eq:secondary capacity}) by substituting
 $p_i$'s that are solutions of (P2) with different values of $C_{\delta}$,
while the corresponding $C_p$'s are obtained from (\ref{eq:primary
capacity}).

\item ``AIPC'': The SU employs the conventional power-control policy
(\ref{eq:optimal power IPC}). $C_s$'s are obtained from
(\ref{eq:secondary capacity})  by substituting
 $p_i$'s that are solutions of (P1) with different values of $\Gamma$, while the corresponding $C_p$'s are obtained from
(\ref{eq:primary capacity}).

\item ``AIPC, Lower Bound'': The SU employs the AIPC-based power-control policy
(\ref{eq:optimal power IPC}). $C_s$ is obtained same as that in the
second case, while for a given value of $\Gamma$, $C_p$ is obtained
as $C_p^{\max}-\log(1+\Gamma)$. Note that $\log(1+\Gamma)$ is shown
in Theorem \ref{theorem} to be a capacity loss upper bound for the
PU and, thus, $C_p$ in this case corresponds to a PU capacity lower
bound.

\item ``MAC, Upper Bound'': An auxiliary 2-user fading Gaussian
multiple-access channel (MAC) is considered here to provide the
capacity upper bounds for the PU and SU. In this auxiliary MAC, all
the channels are the same as those given in Fig. \ref{fig:system
model} except that PU-Rx and SU-Rx are assumed to be collocated such
that the received signals from PU-Tx and SU-Tx can be jointly
processed. Let $\mv{h}_p(i)=[\tilde{\mv{f}}(i) ~
\tilde{\mv{o}}(i)]^T$ and $\mv{h}_s(i)=[\tilde{\mv{g}}(i) ~
\tilde{\mv{e}}(i)]^T$. Considering the auxiliary MAC, for a given PU
power-control policy, $\mathcal{P}_p(f_i)$, it can be shown that the
upper bounds on the PU and SU achievable rates belong to the
following set \cite{Tse}
\begin{align}
\bigcup_{\mathcal{P}_s(h_i,g_i,f_i):\mathbb{E}[p_i]\leq
P}&\bigg\{(C_p,C_s):
C_p\leq\mathbb{E}\left[\log(1+q_i\|\mv{h}_p(i)\|^2)\right],
C_s\leq\mathbb{E}\left[\log(1+p_i\|\mv{h}_s(i)\|^2)\right],
\nonumber
\\ & C_p+C_s\leq
\mathbb{E}\left[\log\left|\mv{I}+q_i\mv{h}_p(i)\mv{h}_p^H(i)+p_i\mv{h}_s(i)\mv{h}_s^H(i)\right|\right]\bigg\}
\end{align}
which can be efficiently computed by applying the methods given in
\cite{Mohseni} and the details are thus omitted here for brevity.
\end{itemize}

Fig. \ref{fig:capacity region CP} and Fig. \ref{fig:capacity region
WF} show the achievable PU and SU ergodic capacities for the
aforementioned cases when the PU power-control policy is the CP
policy in (\ref{eq:CP}) and the WF policy in (\ref{eq:WF}),
respectively. It is observed that in both figures, the capacity
gains by the proposed SU power-control policy using the PCLC are
fairly substantial over the conventional policy using the AIPC. For
example, when the PU capacity loss resulted from the SU transmission
is 5\%, i.e., $C_p=0.95\cdot C_p^{\rm max}$, the SU capacity gain by
the proposed policy over the conventional policy is around 28\% in
the CP case, and 50\% in the WF case. Another interesting
observation is that for the proposed SU power control, the SU
capacity reaches its minimum value of zero in the CP case when the
PU capacity attains its maximum value, $C_p^{\max}$, while in the WF
case the SU capacity has a non-zero value at $C_p^{\max}$. In
general, capacity gains of the proposed SU power control are more
significant in the case of WF over CP PU power control. This is
because the WF policy results in variable PU transmit powers based
on the PU CSI and thus makes the proposed SU power control that is
designed to exploit the PU CSI more beneficial.

\section{Concluding Remarks} \label{sec:conclusion}

In this paper, we studied the fundamental capacity limits for
spectrum sharing based cognitive radio networks over fading
channels. In contrast to the conventional power-control policy for
the SU that applies the interference-power or
interference-temperature constraint at the PU receiver to protect
the PU transmission, this paper proposed a new policy based on the
constraint that limits the maximum permissible PU ergodic capacity
loss resulted from the SU transmission. This new policy is more
directly related to the PU transmission than the conventional one by
exploiting the PU CSI. It was verified by simulation that the
proposed policy can lead to substantial capacity gains for both the
PU and SU over the conventional policy.

Many extensions of this work are possible. First, this paper
considers the ergodic capacity as the performance limits for both PU
and SU, while similar results can be obtained for non-ergodic PU and
SU channels where the outage capacity should be is a more
appropriate measure. Second, results in this paper can also be
extended to the cases with imperfect/quantized PU CSI. Last, the
proposed scheme in this paper is applicable to the general parallel
Gaussian channel with sufficiently large number of sub-channels,
e.g., the broadband channel that is decomposable into a large number
of parallel narrow-band channels via multi-carrier modulation and
demodulation.

\newpage

\begin{figure}
\centering{
 \epsfxsize=6in
    \leavevmode{\epsfbox{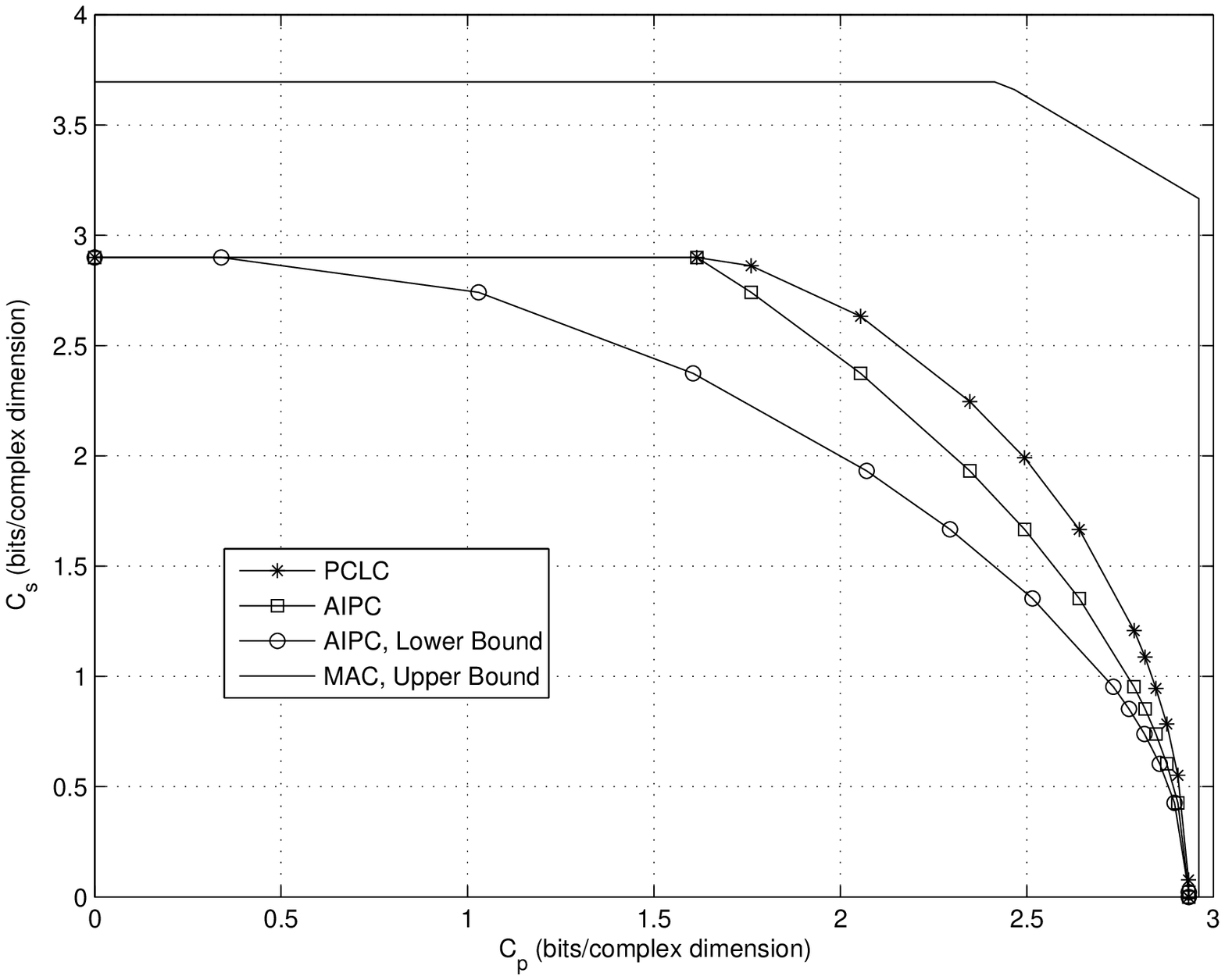}} }
\caption{Achievable rates of PU and SU when the PU power-control
policy is the CP policy given by (\ref{eq:CP}).}\label{fig:capacity
region CP}
\end{figure}

\begin{figure}
\centering{
 \epsfxsize=6in
    \leavevmode{\epsfbox{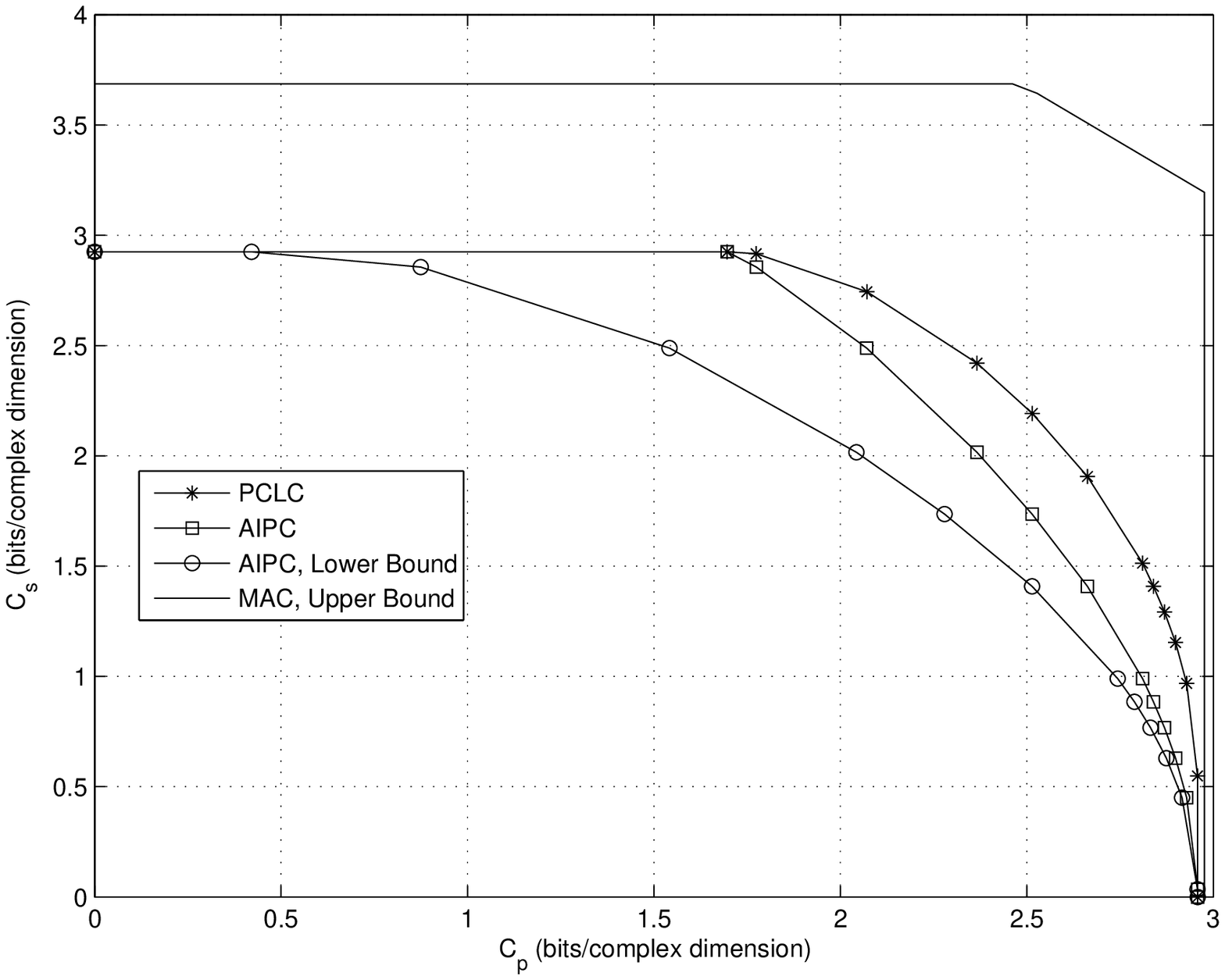}} }
\caption{Achievable rates of PU and SU when the PU power-control
policy is the WF policy given by (\ref{eq:WF}).}\label{fig:capacity
region WF}
\end{figure}

\end{document}